\documentclass{elsart}
\usepackage{latexsym}
\usepackage{epsfig}
\usepackage{natbib}

\def\url#1{#1} 

\begin{document}

\begin{frontmatter}

\title{Magnetic Coupling of a Rotating Black Hole with
Advection-Dominated Accretion Flows}

\author {Yong-Chun Ye, Ding-Xiong Wang\thanksref{email}}
\address{Department of Physics, Huazhong University of Science
and Technology, Wuhan, 430074, People's Republic of China } \and
\author{Ren-Yi Ma}
\address{ Shanghai Astronomical Observatory, Chinese Academy of
Sciences, Shanghai, 200030, People's Republic of China}

      \thanks[email]{E-mail: dxwang@hust.edu.cn}

\begin{abstract}
A model of magnetic coupling (MC) of a rotating black hole (BH)
with advection-dominated accretion flow (MCADAF) is proposed. It
turns out that MCADAF providers a natural explanation for the
transition radius between ADAF and SSD, and could be used to
interpret the highest luminosity of GX 339-4 in hard-state. A very
steep emissivity index can be produced in the innermost part of
the MCADAF, which is consistent with the recent XMM-Newton
observations of the nearby bright Seyfert 1 galaxy MCG-6-30-15 and
with two X-ray binaries (XRBs): XTE J1655-500 and GX 339-4. In
addition, we estimate the BH spins in Seyfert 1 galaxy MCG-6-30-15
and in the two XRBs based on this model.

\end{abstract}


\end{frontmatter}

\section{INTRODUCTION}

More than three decades ago, the standard thin disk was proposed by
Shakura {\&} Sunyayev (1973, hereafter SSD), which is regarded as
the milestone of black hole (BH) accretion theory. However, the
effective radiation temperature is very low in SSD, being obviously
not consistent with the observations of the hard X-ray and gamma-ray
radiation in X-ray binaries (XRBs). In order to overcome this
shortcoming, some authors proposed the cooling-dominated hot disk
(Shapiro, Lightman {\&} Eardley 1976, hereafter SLE). A vital
disadvantage with SLE lies in thermal instability. Not long ago,
advection-dominated accretion flow (hereafter ADAF) was suggested by
some authors (Narayan {\&} Yi 1994, hereafter NY94; 1995a; 1995b;
Abramowicz 1995; Esin 1997; Manmoto 1997; Gammie and Popham 1998;
Popham and Gammie 1998; Yuan 1999). The ADAF model succeeds in
explaining the observed spectra of some sources, such as Sgr A*
(Narayan 1995c), NGC4258 (Lasota 1996) and GRO J1655-40 (Lasota
1998). Recently, Yuan (2001 hereafter Y01; 2003) extended the ADAF
model and put forward a new two-temperature hot branch of
equilibrium solutions, i.e., the luminous hot accretion flow
(hereafter LHAF). It is argued in Y01 that LHAF is much more
luminous than ADAF as the entropy of the accreting mater supply the
radiation as well as the viscous dissipation.

\quad\quad Yuan (2006, hereafter Y06) pointed out that the hard
states at their highest luminosities of some XRBs could not be
explained by ADAF, while LHAF can produce the highest luminosity in
the hard state of XTE J1550-564. However, as argued in Y06 that
neither ADAF nor LHAF could be used to explain the highest
luminosity of GX 339-4 (Done {\&} Gierlinski 2003; Zdziarski et al.
2004), which is a great challenge to LHAF. Probably, an efficient
way of increasing the luminosity in hard-state is the magnetic
coupling (MC) process, which is an important mechanism for
extracting energy from a rotating BH (Blandford 1999; Li {\&}
Paczynski 2000; Li 2002a, hereafter L02; Wang et al. 2002, 2003,
hereafter W02, W03, respectively). However, the MC has been
formulated in the context of a relativistic thin disk (henceforth
MCTHIN, Page {\&} Thorne 1974; L02; W03), in which both the
gravitational energy of the accreting matter and the rotational
energy transferred into the disk are assumed to be radiated away in
black-body spectrum. Thus MCTHIN is obviously not consistent with
observed hard power-law X-ray spectra.

\quad\quad To solve the above puzzle in LHAF we intend to
incorporate the MC with ADAF (henceforth MCADAF). It turns out that
the steep emissivity index in Seyfert 1 galaxy MCG-6-30-15 and the
two XRBs: XTE J1655-500 and GX 339-4 could be interpreted based on
MCADAF, and our fittings are consistent with the low BH spins in
these sources, and that the observation of the highest hard-state
luminosity of GX 339-4 could be explained. Furthermore, a transition
radius between ADAF and SSD could be given naturally in MCADAF with
the hard X-rays spectra produced in the MC region.

\quad\quad This paper is organized as follows. In $\S$ 2 MCADAF is
outlined, and the expressions of the MC power and torque in MCADAF
are derived. In $\S$ 3, the MC effects on disk radiation and the
emissivity index are discussed. It is found that the emissivity
index is consistent with the recent \textit{XMM-Newton} observations
of the nearby bright Seyfert 1 galaxy MCG-6-30-15 and the two XRBs:
XTE J1655-500 and GX 339-4 for the low BH spin. Finally, in $\S$ 4,
we summarize our main results. Throughout this paper the geometric
units $G = c = 1$ are used.

\section{DESCRIPTION OF MCADAF}

\quad\quad In the previous MC model the disk is regarded as a thin
disk, in which the rotational energy transferred from the BH to the
inner disk is assumed to radiate away as the black body spectrum. In
MCADAF, however, this assumption is changed to overcome the
shortcomings of MCTHIN. It is assumed that only a fraction of energy
transferred in the inner disk is radiated, and the rest energy
deposited in the accreting matter as entropy is delivered to the BH
as ADAF. Thus we have the magnetic field configuration of MCADAF as
shown in Figure 1, in which ADAF is located in the MC region with
closed field lines penetrating the accretion flow, and SSD is
located outside the MC region. This configuration is significantly
different from that given in W02 for MCTHIN, and a transition radius
between ADAF and SSD is given naturally in MCADAF.


\begin{figure*}
\vspace{0.5cm}
\begin{center}
\includegraphics[width=10cm]{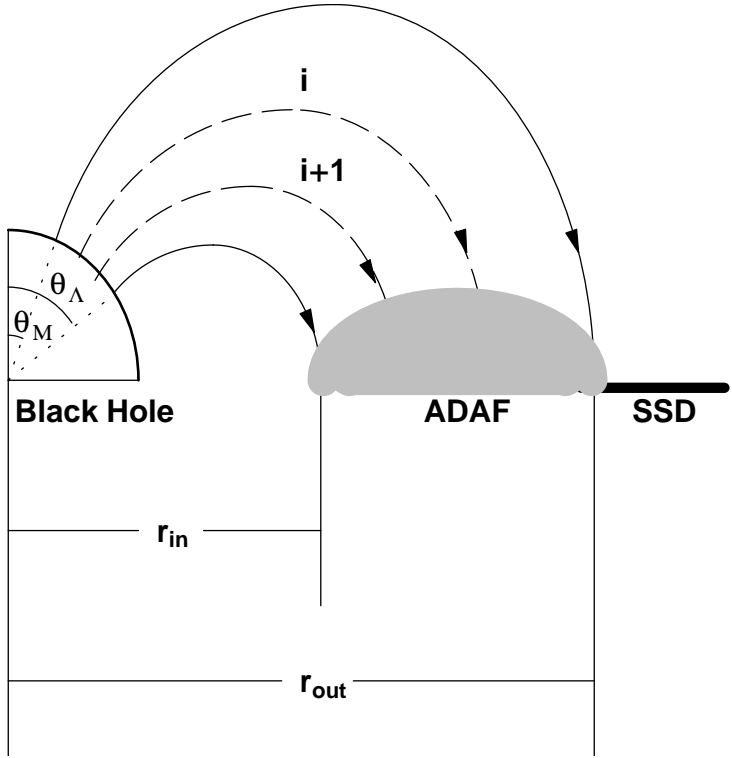}
\caption{The poloidal magnetic field for MCADAF, in which ADAF and
SSD are located inside and outside of $r_{out} $,
respectively.}\label{fig1}
\end{center}
\end{figure*}

\quad\quad In order to facilitate the discussion on MCADAF in an
analytic way, we make the following assumptions.

\quad\quad (i) ADAF is both stable and perfectly conducting, and the
closed magnetic field lines are frozen in the accretion flow.

\quad\quad (ii) The magnetic field is assumed to be weak, and the
instabilities of the magnetic field are ignored. The magnetosphere
is stationary, axisymmetric and force-free outside the BH and ADAF.

\quad\quad Following Armitage {\&} Natarajan (1999), the total
pressure $P$ in the inner part of MCADAF is given by

\begin{equation}
\label{eq1} P = \frac{\sqrt 2 \dot {M}}{12\pi \alpha _{_S} }(5 +
2{\varepsilon }')^{1 / 2}(M / r^5)^{1 / 2},
\end{equation}

\noindent
where the magnetic pressure $P_{mag} $ is related to the total pressure $P$
by


\begin{equation}
\label{eq2} P_{mag} = B_D^2 / 8\pi = (1 - \beta _S )P.
\end{equation}

\quad\quad In equations (\ref{eq1}) and (2) $\alpha _{_S} $ is the
viscosity parameter, $\beta _S $ is the ratio of gas pressure to
total pressure, and ${\varepsilon }'$ is defined as ${\varepsilon }'
= \varepsilon / f = \frac{5 / 3 - \gamma }{f(\gamma - 1)}$ (NY94).
The parameter $f$ measures the degree to which the flow is
advection-dominated, and $\gamma = \frac{8 - 3\beta _S }{6 - 3\beta
_S }$ is the ratio of specific heats (Esin 1997). Thus we have

\begin{equation}
\label{eq3} {\varepsilon }' = \frac{1 - \beta _S }{f}.
\end{equation}

\quad\quad Combining equations (\ref{eq1}) and (2), we have the
expression for the magnetic field in the inner part of the MCADAF as
follows,

\begin{equation}
\label{eq4} B_D = \sqrt {(2 \sqrt {2} / 3)(5 + 2{\varepsilon }')^{1
/ 2} \alpha _{_S} ^{ - 1} (1 - \beta _S ) M^{1/2} \dot {M} r^{-5/2}
} .
\end{equation}

\quad\quad Narayan et al. (1998) pointed out that the inner edge
radius of an ADAF is located between $r_{mb} $ and $r_{ms} $, i.e.,

\begin{equation}
\label{eq5} r_{mb} \le r_{in} \le r_{ms} .
\end{equation}

\noindent
where $r_{mb} $ and $r_{ms} $ are the radii of the innermost bound circular
orbit and the last stable circular orbit, respectively. The radius $r_{in} $
can be expressed by

\begin{equation}
\label{eq6} \left\{ {\begin{array}{l}
 r_{in} \equiv M\chi _{in}^2 , \\
 \chi _{in} = \chi _{mb} + \lambda (\chi _{ms} - \chi _{mb} ),
 \end{array}} \right.
\end{equation}

\noindent where $\chi _{mb} \equiv \sqrt {{r_{mb} } \mathord{\left/
{\vphantom {{r_{mb} } M}} \right. \kern-\nulldelimiterspace} M} $
and $\chi _{ms} \equiv \sqrt {{r_{ms} } \mathord{\left/ {\vphantom
{{r_{ms} } M}} \right. \kern-\nulldelimiterspace} M} $. The
parameter $\lambda $ is used to adjust the position of the inner
edge of the ADAF, and $\lambda = 0.5$ is adopted in calculations.
Thus equation (\ref{eq4}) can be expressed by

\begin{equation}
\label{eq7} B_D \approx B_0 C \xi ^{ - 5 / 4} ,
\end{equation}

\begin{equation}
\label{eq8} B_0 = M^{ - 1}\dot {M}^{1 / 2} = 1.38\times
10^9m_{_{BH}}^{ - 1 / 2} \dot {m}^{1 / 2}Gauss,
\end{equation}

\begin{equation}
\label{eq9} C = (2\sqrt 2 / 3)^{1 / 2}(5 + 2{\varepsilon }')^{1 /
4}\alpha _{_S} ^{ - 1 / 2}(1 - \beta _S )^{1 / 2}\chi _{in}^{ - 5 /
2} ,
\end{equation}


\begin{figure}
\vspace{0.5cm}
\begin{center}
{\includegraphics[width=7.5cm]{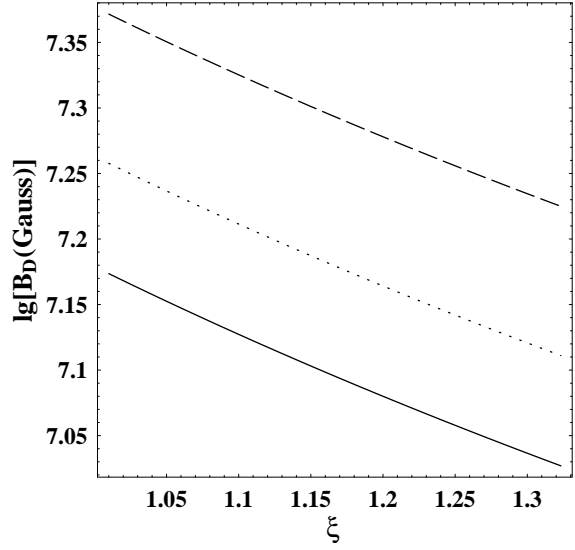}
 \centerline{\quad\quad\quad(a)}
 \includegraphics[width=7.5cm]{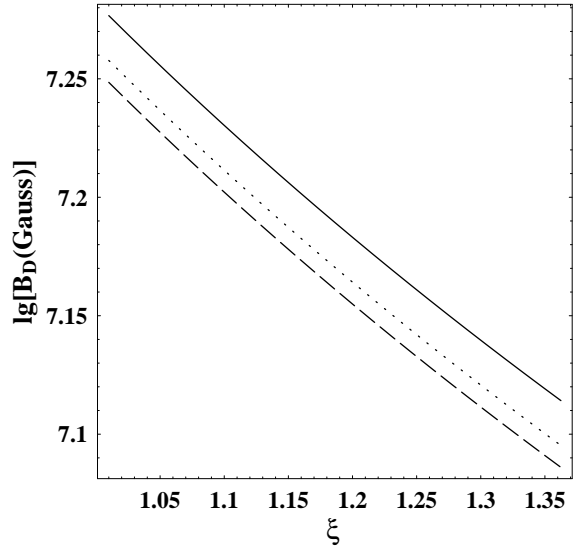}
 \centerline{\quad\quad\quad(b)}}
 \caption{The curves of $\lg [B_D (Gauss)]$ in MCADAF
versus $\xi $ with $m_{_{BH}} = 10$ for (a) $f$=0.5 and $a_\ast $=
0.3, 0.5 and 0.7 in solid, dotted and dashed lines, respectively,
and for (b) $a_\ast $=0.5 and $f$= 0.3, 0.5 and 0.7 in solid, dotted
and dashed lines, respectively.}\label{fig2}
\end{center}
\end{figure}

\noindent where $\xi \equiv r \mathord{\left/ {\vphantom {r {r_{in}
}}} \right. \kern-\nulldelimiterspace} {r_{in} }$ is a dimensionless
radial parameter, $m_{_{BH}}$ is the BH mass in the units of one
solar mass, and $\dot {m} \equiv \dot {M} \mathord{\left/ {\vphantom
{\dot {M} {\left( {{L_E } \mathord{\left/ {\vphantom {{L_E } {c^2}}}
\right. \kern-\nulldelimiterspace} {c^2}} \right)}}} \right.
\kern-\nulldelimiterspace} {\left( {{L_E } \mathord{\left/
{\vphantom {{L_E } {c^2}}} \right. \kern-\nulldelimiterspace} {c^2}}
\right)}$ is the accretion rate in terms of the Eddington
luminosity.

\quad\quad Incorporating equations (\ref{eq3})---(9), we can
calculate the magnetic field $B_D $ in MCADAF versus $\xi $ with the
given $a_\ast $ and $f$ as shown in Figure 2. In the following
calculations, $\dot {m} = 0.01$, $\alpha _{_S} = 0.3$, $\beta _S =
0.5$ are adopted as given by Narayan et al. (1998).

\quad\quad The magnetic field in MCTHIN varying with the radial
parameter $\xi $ is given in W03 as follows,

\begin{equation}
\label{eq10} B_D^p = B_H^p \left[ {{r_{_H} } \mathord{\left/
{\vphantom {{r_{_H} } {\varpi _{_D} \left( {r_{ms} } \right)}}}
\right. \kern-\nulldelimiterspace} {\varpi _{_D} \left( {r_{ms} }
\right)}} \right]\xi ^{ - n}.
\end{equation}

\quad\quad Incorporating equation (\ref{eq10}) with some relations
given in W03, we have the curves of the magnetic field $B_D^p $
versus $\xi $ with given $a_\ast $ and $n$ in MCTHIN as shown in
Figure 3.


\begin{figure}
\vspace{0.5cm}
\begin{center}
\includegraphics[width=10cm]{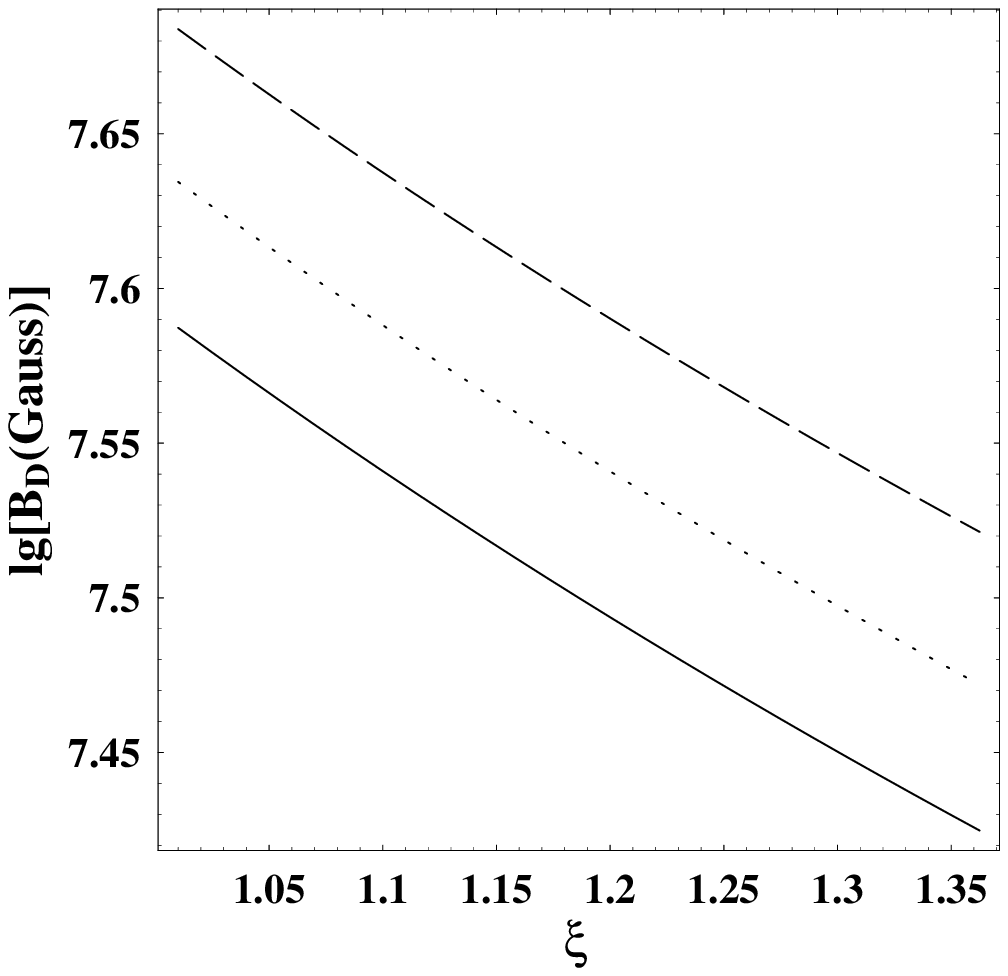}
\caption{The curves of magnetic field $\lg [B_D (Gauss)]$ in MCTHIN
versus $\xi $ with $n = 5 / 4$ and $a_\ast $= 0.3, 0.5 and 0.7 in
solid, dotted and dashed lines, respectively.}\label{fig3}
\end{center}
\end{figure}

\quad\quad Inspecting Figures 2 and 3, we find that the magnetic
field in MCADAF is weaker than that in MCTHIN.

\quad\quad The mapping relation between the angular coordinate on
the horizon and the radial coordinate in MCADAF can be derived based
on the conservation of magnetic flux in an analogous way given in
W03, and we have

\begin{equation}
\label{eq11} \sin \theta d\theta = - \mbox{R}\left( {a_ * ,\xi }
\right)d\xi ,
\end{equation}

\noindent
where

\begin{equation}
\label{eq12} \mbox{R}\left( {a_ * ,\xi } \right) = \frac{\xi ^{ - 1
/ 4}\chi _{in}^2 \sqrt {1 + a_ * ^2 \chi _{in}^{ - 4} \xi ^{ - 2} +
2a_ * ^2 \chi _{in}^{ - 6} \xi ^{ - 3}} }{2\sqrt {1 + \chi _{in}^{ -
4} a_ * ^2 + 2\chi _{in}^{ - 6} a_ * ^2 } \sqrt {1 - 2\chi _{in}^{ -
2} \xi ^{ - 1} + a_ * ^2 \chi _{in}^{ - 4} \xi ^{ - 2}} }.
\end{equation}

Integrating equation (\ref{eq11}) and setting $\xi = \xi _{in} = 1$
at $\theta _L = 0.45\pi $, we have

\begin{equation}
\label{eq13} \cos \theta _M - \cos \theta _L = \int_1^{\xi _{out} }
{\mbox{R}\left( {a_ * ,\xi } \right)} d\xi .
\end{equation}

\quad\quad The parameter $\xi _{out} \equiv {r_{out} }
\mathord{\left/ {\vphantom {{r_{out} } {r_{in} }}} \right.
\kern-\nulldelimiterspace} {r_{in} }$ in equation (\ref{eq13})
represents the outer radius of the MC region, and we have the curves
of $\xi _{out} $ versus $a_ * $ with different values of $\theta _M
$ as shown in Figure 4.


\begin{figure}
\vspace{0.5cm}
\begin{center}
\includegraphics[width=10cm]{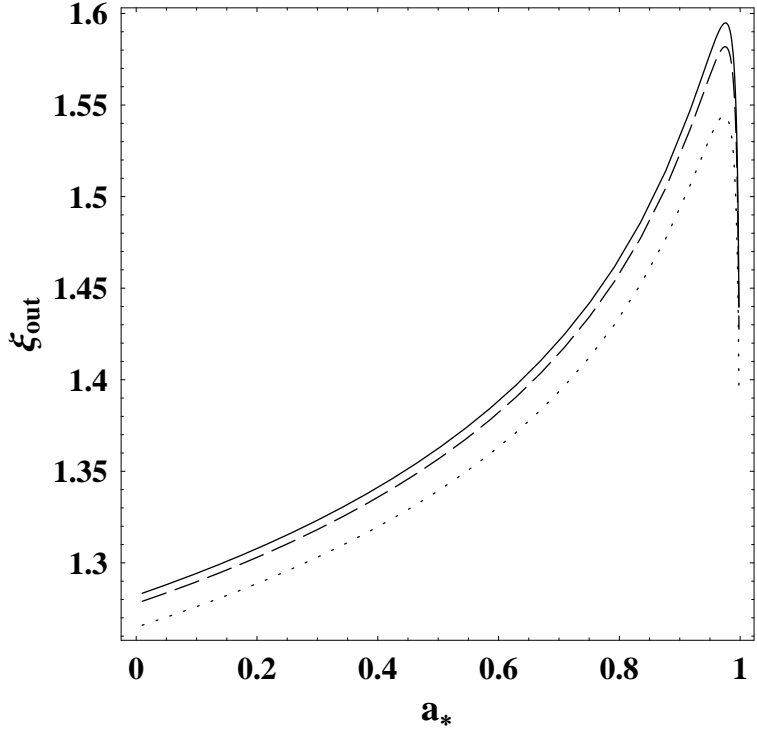}
\caption{The outer radius of the MC region, $\xi _{out} $, versus
$a_\ast $ with $\theta _M $= 0, $0.05\pi $ and $0.1\pi $ in solid,
dotted and dashed lines respectively.}\label{fig4}
\end{center}
\end{figure}

\quad\quad From Figure 4 we find that the variation of $\xi _{out} $
is insensitive to the value of $\theta _M $, and the MC region in
MCADAF is restricted to a very small range, $1 \le \xi \le (\xi
_{out} )_{\max } = \mbox{1.595}$. This result implies that the
transition radius in MCADAF is very small.

\quad\quad Also analogous to W02 we can derive the expression for
the MC power and torque in MCADAF by using an equivalent circuit in
BH magnetosphere (Macdonald {\&} Thorne 1982), and the torque and
power exerted on the MCADAF between two adjacent magnetic surfaces
are given by

\begin{equation}
\label{eq14} \Delta T_{MC} = \varpi B_H I\Delta l = \left( {{\Delta
\Psi } \mathord{\left/ {\vphantom {{\Delta \Psi } {2\pi }}} \right.
\kern-\nulldelimiterspace} {2\pi }} \right)^2{\left( {\Omega _H -
\Omega _D } \right)} \mathord{\left/ {\vphantom {{\left( {\Omega _H
- \Omega _D } \right)} {\Delta Z_H }}} \right.
\kern-\nulldelimiterspace} {\Delta Z_H },
\end{equation}

\begin{equation}
\label{eq15} \Delta P_{MC} = \Omega _F \Delta T = \left( {{\Delta
\Psi } \mathord{\left/ {\vphantom {{\Delta \Psi } {2\pi }}} \right.
\kern-\nulldelimiterspace} {2\pi }} \right)^2{\Omega _F \left(
{\Omega _H - \Omega _D } \right)} \mathord{\left/ {\vphantom
{{\Omega _F \left( {\Omega _H - \Omega _D } \right)} {\Delta Z_H }}}
\right. \kern-\nulldelimiterspace} {\Delta Z_H },
\end{equation}

\noindent
where

\begin{equation}
\label{eq16} \Delta Z_H = R_{H} {\Delta l} \mathord{\left/
{\vphantom {{\Delta l} {\left( {2\pi \varpi } \right)}}} \right.
\kern-\nulldelimiterspace} {\left( {2\pi \varpi } \right)} = {2\rho
\Delta \theta } \mathord{\left/ {\vphantom {{2\rho \Delta \theta }
\varpi }} \right. \kern-\nulldelimiterspace} \varpi .
\end{equation}

\quad\quad Incorporating the expression for angular velocity given
in NY94, we express the angular velocity $\Omega _D $ in MCADAF as
follows,

\begin{equation}
\label{eq17} \Omega _D \approx \left[ {{2{\varepsilon }'}
\mathord{\left/ {\vphantom {{2{\varepsilon }'} {\left( {5 +
2{\varepsilon }'} \right)}}} \right. \kern-\nulldelimiterspace}
{\left( {5 + 2{\varepsilon }'} \right)}} \right]^{1 / 2}\Omega _K ,
\quad \Omega _K = \frac{1}{M\left( {a_\ast + \xi ^{3 \mathord{\left/
{\vphantom {3 2}} \right. \kern-\nulldelimiterspace} 2}\chi _{in}^3
} \right)}.
\end{equation}

\quad\quad Integrating equations (\ref{eq14}) and (\ref{eq15}) over
the angular coordinate from $\theta _M $ to $\theta _L = 0.45\pi $,
we obtain the total MC power and torque as follows.

\begin{equation}
\label{eq18} {T_{MC} } \mathord{\left/ {\vphantom {{T_{MC} } {T_0
}}} \right. \kern-\nulldelimiterspace} {T_0 } = \frac{4a_ * C^2\chi
_{in}^4 (1 + \chi _{in}^{ - 4} a_ * ^2 + 2\chi _{in}^{ - 6} a_ * ^2
)}{1 + q}\int_{\theta _M }^{\theta _L } {\frac{\left( {1 - \beta }
\right)\sin ^3\theta d\theta }{2 - \left( {1 - q} \right)\sin
^2\theta }} ,
\end{equation}

\begin{equation}
\label{eq19} {P_{MC} } \mathord{\left/ {\vphantom {{P_{MC} } {P_0
}}} \right. \kern-\nulldelimiterspace} {P_0 } = \frac{2a_ * ^2
C^2\chi _{in}^4 (1 + \chi _{in}^{ - 4} a_ * ^2 + 2\chi _{in}^{ - 6}
a_ * ^2 )}{(1 + q)^2}\int_{\theta _M }^{\theta _L } {\frac{\beta
\left( {1 - \beta } \right)\sin ^3\theta d\theta }{2 - \left( {1 -
q} \right)\sin ^2\theta }} ,
\end{equation}

\noindent
where

\begin{equation}
\label{eq20} \beta \equiv {\Omega _D } \mathord{\left/ {\vphantom
{{\Omega _D } {\Omega _H }}} \right. \kern-\nulldelimiterspace}
{\Omega _H } = \frac{2(1 + q)\left[ {{2{\varepsilon }'}
\mathord{\left/ {\vphantom {{2{\varepsilon }'} {\left( {5 +
2{\varepsilon }'} \right)}}} \right. \kern-\nulldelimiterspace}
{\left( {5 + 2{\varepsilon }'} \right)}} \right]^{1 / 2}}{a_\ast
(a_\ast + \xi ^{3 \mathord{\left/ {\vphantom {3 2}} \right.
\kern-\nulldelimiterspace} 2}\chi _{in}^3 )},
\end{equation}

\begin{equation}
\label{eq21} \left. {\begin{array}{l}
 P_0 = \left\langle {B_0^2 } \right\rangle M^2 \approx m_{_{BH}} \dot {m}\times
1.25\times 10^{36}erg \cdot s^{ - 1} \\
 T_0 = \left\langle {B_0^2 } \right\rangle M^3 \approx m_{{BH}}^2 \dot
{m}\times 5.52\times 10^{29}g \cdot cm^2 \cdot s^{ - 2} \\
 \end{array}} \right\} .
\end{equation}

\quad\quad The co-rotation radius $r_c $ can be determined from the
equation (\ref{eq20}) by setting $\beta = 1$, which indicates the
place at ADAF with $\Omega _D = \Omega _H $. Incorporating equations
(\ref{eq12}), (\ref{eq13}) and (\ref{eq20}), we have the curves of
dimensionless co-rotation radius $\xi _c \equiv {r_c }
\mathord{\left/ {\vphantom {{r_c } {r_{in} }}} \right.
\kern-\nulldelimiterspace} {r_{in} }$ and $\xi _{out} $ versus $a_ *
$ for the different values of $f$ with $\theta _M = 0$ as shown in
Figure 5.


\begin{figure}
\vspace{0.5cm}
\begin{center}
\includegraphics[width=9.5cm]{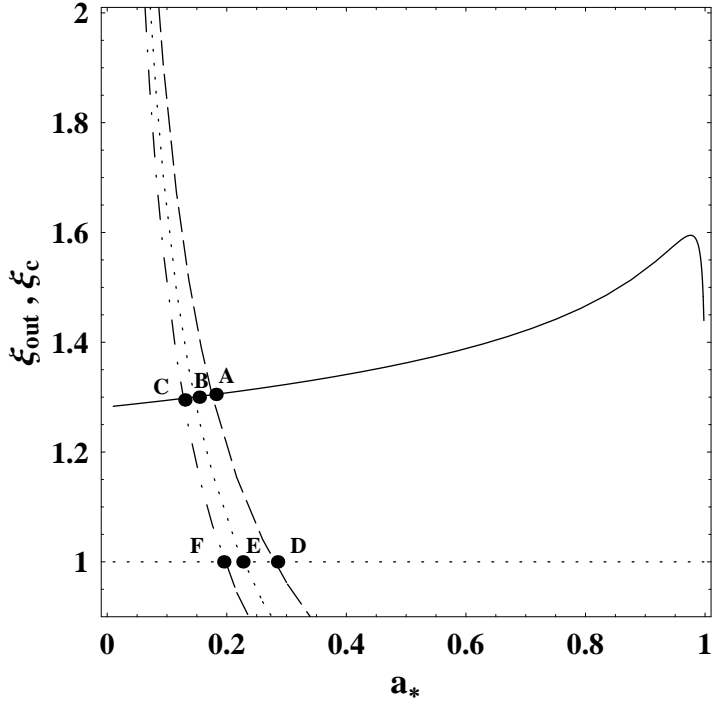}
\caption{The curves of $\xi _{out} $ (solid line) and $\xi _c $
versus $a_ * $ with $f$=0.3, 0.5, 0.7 in dashed, dotted and
dot-dashed lines, respectively.}\label{fig5}
\end{center}
\end{figure}

\quad\quad Inspecting Figure 5, we find that each curve of $\xi _c $
decreases monotonically with $a_ * $, and it intersects with the
curves of $\xi _{out} $ and $\xi _{in} = 1$, respectively. These
intersections imply that the co-rotation radius $r_c $ is located at
the outer and inner boundary of the MC region for the BH spin equal
to $a_ * ^{out} $ and $a_ * ^{in} $, respectively.

\quad\quad The MC region is divided into two parts by $\xi _c $,
i.e., the inner MC region for $1 < \xi < \xi _c $ and the outer MC
region for $\xi _c < \xi < \xi _{out} $ with $a_\ast ^{out} < a_\ast
< a_\ast ^{in} $. Thus the energy and angular momentum are
transferred by the MC from the BH into the outer MC region with
$\Omega _D < \Omega _H $, while the transfer direction reverses for
the inner MC region with $\Omega _D > \Omega _H $. The correlation
of the BH spin with the transfer direction is given as shown in
Table 1.

\begin{table*}
\begin{center}
 \caption{The correlation of the BH spin with the transfer direction
of energy and angular momentum}
\end{center}
\begin{tabular}
{|p{14pt}|p{20pt}|p{23pt}|p{85pt}|p{85pt}|p{85pt}|} \hline
\multicolumn{3}{|p{57pt}|}{Parameters} &
\multicolumn{3}{|p{255pt}|}{Transfer direction of energy and angular momentum}  \\
\hline \raisebox{-1.50ex}[0cm][0cm]{$f$}&
\raisebox{-1.50ex}[0cm][0cm]{$a_\ast ^{in} $}&
\raisebox{-1.50ex}[0cm][0cm]{$a_\ast ^{out} $}&
\multicolumn{2}{|p{171pt}|}{$a_\ast ^{out} < a_\ast < a_\ast ^{in} $
} &
$a_\ast ^{in} < a_\ast < 0.998$  \\
\cline{4-6}
 &
 &
 &
$1 < \xi < \xi _c $& $\xi _c < \xi < \xi _{out} $&
$\xi _c $ does not exist \\
\hline 0.3& 0.279& 0.175& \raisebox{-3.00ex}[0cm][0cm]{\textbf{from
disk to BH} \par }& \raisebox{-3.00ex}[0cm][0cm]{\textbf{from BH to
disk} \par }&
\raisebox{-3.00ex}[0cm][0cm]{\textbf{from BH to disk} \par } \\
\cline{1-3} 0.5& 0.228& 0.146&
 &
 &
  \\
\cline{1-3} 0.7& 0.196& 0.127&
 &
 &
  \\
\hline
\end{tabular}
 \label{tab1}
\end{table*}

\quad\quad By using equations (\ref{eq11})---(\ref{eq21}), we have
the curves of the MC power and torque versus $a_ * $ for different
values of $f$ as shown in Figure 6. It is found that the MC power
increases monotonically with the increasing BH spin, while the MC
torque increases generally with the BH spin, except the spin
approaching unity. It is noted that both the MC power and torque are
insensitive to the values of the parameter $f$.


\begin{figure}
\vspace{0.5cm}
\begin{center}
{\includegraphics[width=7.5cm]{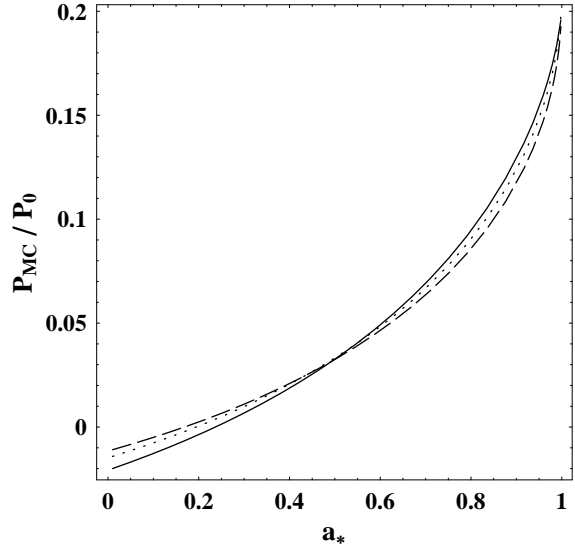}
 \centerline{\quad\quad\quad(a)}
 \includegraphics[width=7.5cm]{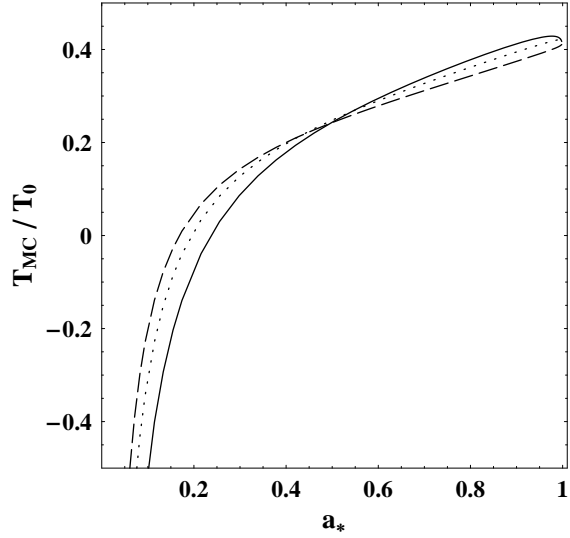}
 \centerline{\quad\quad\quad(b)}}
 \caption{The curves of $P_{MC} / P_0 $ (upper panel) and
$T_{MC} / T_0 $ (lower panel) versus $a_ * $ with $f$=0.3, 0.5, 0.7
in solid, dotted, dashed lines, respectively.}\label{fig6}
\end{center}
\end{figure}

\section{STEEP EMISSIVITY INDEX PRODUCED IN MCADAF}

\quad\quad Based on the MC effects on radiation flux we can
calculate the emissivity index in MCADAF. The energy conservation
equation given in NY94 is

\begin{equation}
\label{eq22} Q^ + - Q^ - = \frac{2\alpha _{_S} \rho c_s^2
r^2H}{\Omega _K }(\frac{d\Omega }{dr})^2 - Q^ - = f\frac{2\alpha
_{_S} \rho c_s^2 r^2H}{\Omega _K }(\frac{d\Omega }{dr})^2,
\end{equation}

\noindent where $Q^ + $ is the energy input per unit area due to
viscous dissipation, and $Q^ - $ is the energy loss due to radiative
cooling. Considering the electromagnetic flux of energy transferred
by the MC process, we modify equation (\ref{eq22}) as follows,

\begin{equation}
\label{eq23} \frac{2\alpha \rho c_s^2 r^2H}{\Omega _K
}(\frac{d\Omega }{dr})^2 + H_{MC} \Omega - Q^ - = f\frac{2\alpha
_{_S} \rho c_s^2 r^2H}{\Omega _K }(\frac{d\Omega }{dr})^2,
\end{equation}

\noindent
where $H_{MC} $ is the angular momentum flux transferred in MCADAF, which is
related to $T_{MC} $ by (Li 2002b, W03)

\begin{equation}
\label{eq24}  \begin{array}{l} {\partial T_{MC} } \mathord{\left/
{\vphantom {{\partial T_{MC} } {\partial r}}} \right.
\kern-\nulldelimiterspace} {\partial r} = 4\pi rH_{MC} \\
\quad\quad = - \left( {\frac{4 T_0 a_ * \left( {1 - \beta }
\right)\sin ^3\theta }{2 - \left( {1 - q} \right)\sin ^2\theta }}
\right)\left[ {\frac{C^2\chi _{in}^4 (1 + \chi _{in}^{ - 4} a_ * ^2
+ 2\chi _{in}^{ - 6} a_ * ^2 )}{1 + q}} \right]\left(
{\frac{\partial \theta }{\partial r}} \right),
\end{array}
\end{equation}

\noindent and ${\partial \theta } \mathord{\left/ {\vphantom
{{\partial \theta } {\partial r}}} \right.
\kern-\nulldelimiterspace} {\partial r}$ can be calculated by using
equation (\ref{eq12}),

\begin{equation}
\label{eq25} {\partial \theta } \mathord{\left/ {\vphantom
{{\partial \theta } {\partial r}}} \right.
\kern-\nulldelimiterspace} {\partial r} = \left( {{\partial \theta }
\mathord{\left/ {\vphantom {{\partial \theta } {\partial \xi }}}
\right. \kern-\nulldelimiterspace} {\partial \xi }} \right)\left(
{{\partial \xi } \mathord{\left/ {\vphantom {{\partial \xi }
{\partial r}}} \right. \kern-\nulldelimiterspace} {\partial r}}
\right) = - \frac{\mbox{R}\left( {a_ * ,\xi } \right)}{r_{in} \sin
\theta }.
\end{equation}

Incorporating equations (24) and (\ref{eq25}), we have

\begin{equation}
\label{eq26} {H\left( {a_ * ,\xi } \right)} \mathord{\left/
{\vphantom {{H\left( {a_ * ,\xi } \right)} {H_0 }}} \right.
\kern-\nulldelimiterspace} {H_0 } = \left\{ {\begin{array}{l}
 A\left( {a_ * ,\xi } \right), 1 < \xi  < \xi _{out}, \\
 0,\mbox{ }\xi > \xi _{\mbox{out}} \mbox{ }, \\
 \end{array}} \right.
\end{equation}

\noindent
where

\begin{equation}
\label{eq27} H_0 = \left\langle {B_0^2 } \right\rangle M = \dot
{m}\times 3.15\times 10^{20}\mbox{ }g \cdot s^{ - 2},
\end{equation}

\begin{equation}
\label{eq28} A\left( {a_ * ,\xi } \right) = \frac{C^2(1 + \chi
_{in}^{ - 4} a_ * ^2 + 2\chi _{in}^{ - 6} a_ * ^2 )}{1 + q}\frac{a_
* \left( {1 - \beta } \right)}{\pi \left[ {2\csc ^2\theta - \left(
{1 - q} \right)} \right]}R\left( {a_ * ,\xi } \right).
\end{equation}

Then we obtain the expression for radiation flux in MCADAF as follows,

\begin{equation}
\label{eq29} F_{total} = Q^ - = H_{MC} \Omega + (1 - f)\frac{2\alpha
_{_S} \rho c_s^2 R^2H}{\Omega _K }(\frac{d\Omega }{dr})^2.
\end{equation}

\quad\quad Combining equations (24)---(\ref{eq29}), we have

\begin{equation}
\label{eq30}  \begin{array}{l}
F_{total} / F_0 = A\left( {a_ * ,\xi
} \right)\frac{\left[ {{2{\varepsilon }'} \mathord{\left/ {\vphantom
{{2{\varepsilon }'} {\left( {5 + 2{\varepsilon }'} \right)}}}
\right. \kern-\nulldelimiterspace} {\left( {5 + 2{\varepsilon }'}
\right)}} \right]^{1 / 2}}{a_\ast + \xi ^{3 \mathord{\left/
{\vphantom {3 2}} \right. \kern-\nulldelimiterspace} 2}\chi _{in}^3
} + \frac{(1 - f)\xi ^2}{3\pi }\left( {\frac{\partial }{\partial \xi
}\left\{ {\frac{\left[ {{2{\varepsilon }'} \mathord{\left/
{\vphantom {{2{\varepsilon }'} {\left( {5 + 2{\varepsilon }'}
\right)}}} \right. \kern-\nulldelimiterspace} {\left( {5 +
2{\varepsilon }'} \right)}} \right]^{1 / 2}}{a_\ast + \xi ^{3
\mathord{\left/ {\vphantom {3 2}} \right. \kern-\nulldelimiterspace}
2}\chi _{in}^3 }} \right\}} \right)^2,
\end{array}
\end{equation}

\quad\quad In equation (\ref{eq30}) we have $F_0 = \left\langle
{B_0^2 } \right\rangle c$, and the first and second terms at RHS
represent the MC effect on radiation flux and the contribution due
to ADAF, respectively.

\quad\quad Following L02 and W03, the emissivity index is defined as

\begin{equation}
\label{eq31} \alpha \equiv - {d\ln F} \mathord{\left/ {\vphantom
{{d\ln F} {d\ln r}}} \right. \kern-\nulldelimiterspace} {d\ln r}.
\end{equation}

\quad\quad By using equations (24)---(\ref{eq31}), we have the
curves of the emissivity index $\alpha $ versus $\lg \left( {r
\mathord{\left/ {\vphantom {r M}} \right. \kern-\nulldelimiterspace}
M} \right)$ for the different values of $a_\ast $ and $f$ as shown
in Figure 7. As shown in the shaded region of Figure 7, we find that
the recent \textit{XMM-Newton} observation of the nearby bright
Seyfert 1 galaxy MCG-6-30-15 (Wilms et al. 2001) can be simulated in
MCADAF, while the index produced by the ADAF is far below the shaded
region as shown by the dot-dashed lines in Figure 7.


\begin{figure}
\vspace{0.5cm}
\begin{center}
{\includegraphics[width=7.5cm]{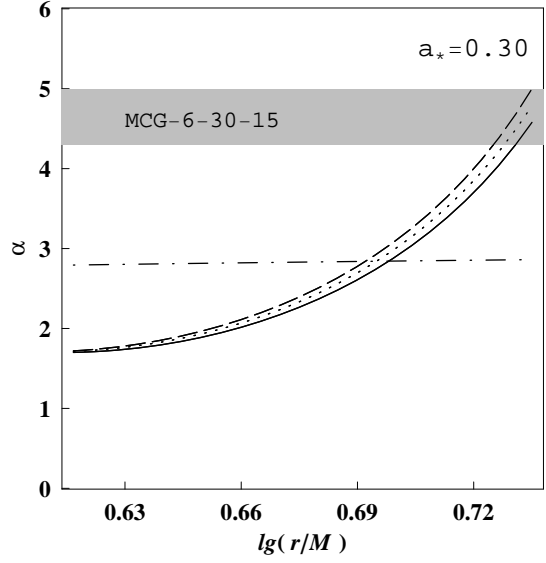}
 \centerline{\quad\quad\quad(a)}
 \includegraphics[width=7.5cm]{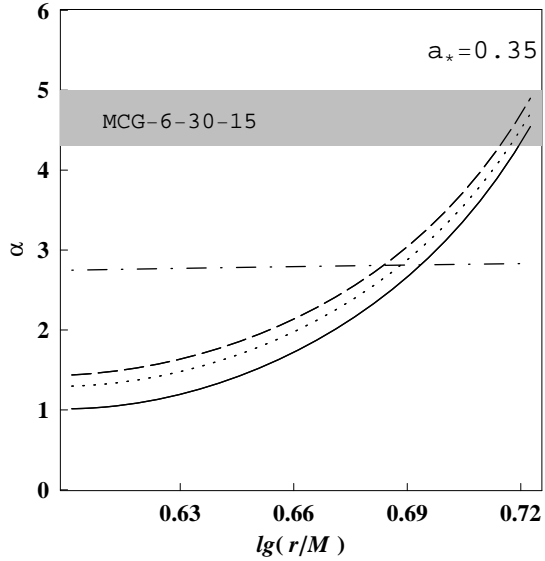}
 \centerline{\quad\quad\quad(b)}}
 \caption{The emissivity index of the Seyfert 1 galaxy MCG-6-30-15
inferred from the observation of \textit{XMM-Newton} is shown in the
shaded region. The emissivity index in MCADAF is plotted versus $\lg
( r / M )$ for $a_\ast $=0.30 and 0.35 in panels (a) and (b),
respectively. The curves are plotted for $f$=0.3, 0.325 and 0.35 in
solid, dotted and dashed lines, respectively. The emissivity index
in ADAF without MC is plotted in dot-dashed line.}\label{fig7}
\end{center}
\end{figure}

\quad\quad Comparing the results in Figure 7 with those obtained in
W03, we find that the emissivity index in MCADAF corresponds to the
spins less than 0.4, which are much lower than those obtained in
MCTHIN with $a_\ast > 0.99$. By analyzing the X-ray spectrum of the
Seyfert 1 galaxy MCG-6-30-15, Dovciak et al. (2004) discussed a
low-spin ($a_\ast = 0.25)$ best-fitting model for the iron line
emission in MCG-6-30-15. This result implied that the central BH
spin may be low. Brenneman1 and Reynolds (2006) pointed out that an
inner emissivity profile requires an inner radius of $r_{inner} =
3.2r_g $ ($r_g \equiv M)$ to fit the iron line emission. These
results are consistent with MCADAF but not with MCTHIN, since the
inner edge of MCADAF is located at $r_{in} < 3.2r_g $ for a low BH
spin, and the inner emissivity profile can be produced by virtue of
the energy transferred in the MC process.

\quad\quad It is shown by the recent observations that the very
steep emissivities are also found in XTE J1650-500 (Miller et al.
2002; Miniutti et al. 2004) and GX 339-4 (Miller et al. 2004). The
simulations for the steep emissivity in these XRBs can be worked out
in the same way as MCG-6-30-15, and the curves of the emissivity
index $\alpha $ versus $\lg \left( {r \mathord{\left/ {\vphantom {r
M}} \right. \kern-\nulldelimiterspace} M} \right)$ for the different
values of $a_\ast $ and $f$ are shown in Figure 8, from which we
find that the BH spins of the two XRBs are also estimated as the low
values in fitting the observed very steep emissivity indexes.


\begin{figure}
\vspace{0.5cm}
\begin{center}
{\includegraphics[width=7.5cm]{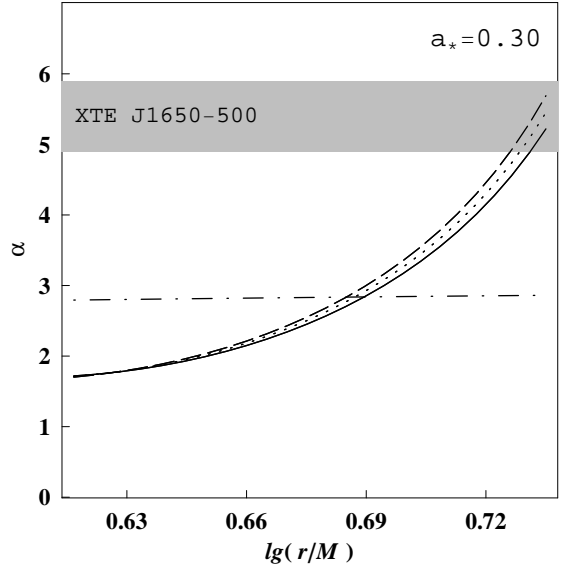}
 \centerline{\quad\quad\quad(a)}
 \includegraphics[width=7.5cm]{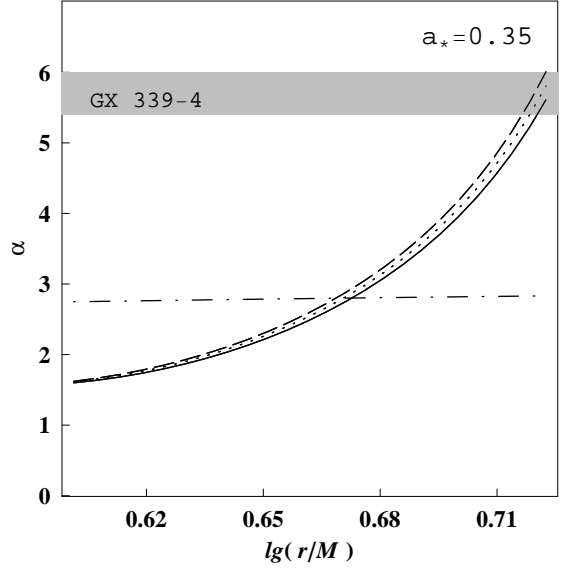}
 \centerline{\quad\quad\quad(b)}}
 \caption{The emissivity indexes of XTE J1650-500 (upper panel) and
GX 339-4 (lower panel) inferred from the observation of
\textit{XMM-Newton} are shown in the shaded region. The emissivity
indexes produced in MCADAF versus $\lg ( r / M ) $ are plotted in
upper panel in solid, dotted and dashed lines for $a_\ast $=0.30
with $f$=0.25, 0.275 and 0.3, respectively. Those are plotted in
lower panel in solid, dotted and dashed lines for $a_\ast $=0.35
with $f$=0.18, 0.2 and 0.22, respectively. The emissivity index
produced in ADAF without MC is plotted in dot-dashed
line.}\label{fig8}
\end{center}
\end{figure}

\section{DISCUSSION}

\quad\quad In this paper, we propose a toy model for
advection-dominated accretion flows around a rotating BH with the
global magnetic field. And the expressions for the extracting power
and torque in the MC process and the MC effects on radiation flux
are derived. By simulating the emissivity index in Seyfert 1 galaxy
MCG-6-30-15 and the two XRBs: XTE J1655-500 and GX 339-4, we find
that the steep emissivity index could be interpreted based on
MCADAF, and our fittings are consistent with the low BH spins in
these sources.

\quad\quad Comparing with MCTHIN and ADAF (or LHAF), the advantage
of MCADAF lies in the following aspects.

\quad\quad (\ref{eq1}) The electrons of high-temperature in the
optically thin plasma are contained in MCADAF, and they are cooled
via synchrotron, bremsstrahlung and inverse Compton processes, being
responsible for producing hard X-rays spectra in a natural way.

\quad\quad (2) MCADAF providers a natural explanation for transition
radius between ADAF and SSD as shown in Figure 1.

\quad\quad (\ref{eq3}) MCADAF could be used to interpret the highest
luminosity of GX 339-4 in hard-state.

\quad\quad Following W02, we have the energy extracting efficiency
in MCADAF as follows:

\begin{equation}
\label{eq32} \eta _{_{MCADAF}} = \eta _{_{ADAF}} + \eta _{_{MC}}
\end{equation}

\noindent where
\begin{equation} \label{eq33}  \left\{ {\begin{array}{l} \eta _{_{ADAF}} = 1 - E_{in} , \\{\eta _{_{MC}} = P_{MC} } \mathord{\left/
{\vphantom {{\eta _{_{MC}} = P_{MC} } \dot {M}}} \right.
\kern-\nulldelimiterspace} \dot {M}, \\E_{in} = {\left( {1 - 2\chi
_{in}^{ - 2} + a_ * \chi _{in}^{ - 3} } \right)} \mathord{\left/
{\vphantom {{\left( {1 - 2\chi _{in}^{ - 2} + a_ * \chi _{in}^{ - 3}
} \right)} {\left( {1 - 3\chi _{in}^{ - 2} + 2a_ * \chi _{in}^{ - 3}
} \right)^{1 / 2}}}} \right. \kern-\nulldelimiterspace} {\left( {1 -
3\chi _{in}^{ - 2} + 2a_ * \chi _{in}^{ - 3} } \right)^{1 / 2}}.
\end{array}} \right.
\end{equation}

\quad\quad By using equations (\ref{eq32}) and (\ref{eq33}), we have
the curves of $\eta _{_{MCADAF}} $ and $\eta _{_{ADAF}} $ versus
$a_\ast $, and the ratio of the former to the latter versus $a_\ast
$ as shown in Figure 9.


\begin{figure}
\vspace{0.5cm}
\begin{center}
{\includegraphics[width=7.5cm]{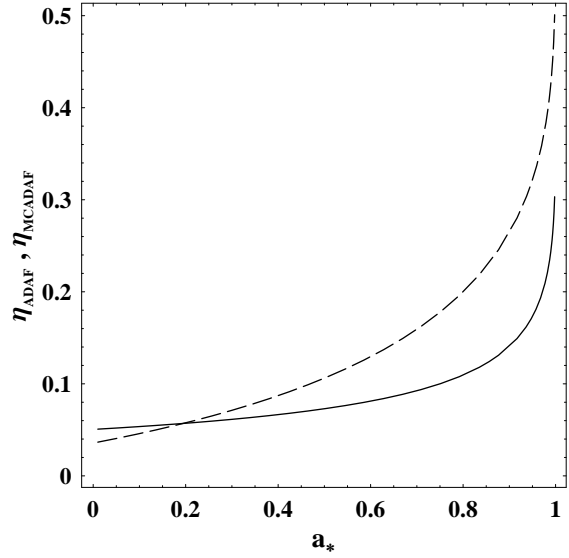}
 \centerline{\quad\quad\quad(a)}
 \includegraphics[width=7.5cm]{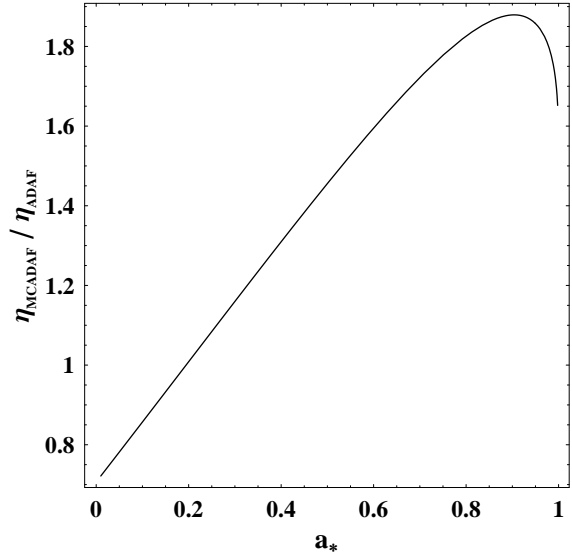}
 \centerline{\quad\quad\quad(b)}}
 \caption{The curves of $\eta _{_{ADAF}} $ (solid line) and
$\eta _{_{MCADAF}} $ (dashed line) versus $a_ * $ with $f$=0.5; (b)
the ratio $\eta _{_{MCADAF}} / \eta _{_{ADAF}}$ versus $a_ * $ with
$f$=0.5.}\label{fig9}
\end{center}
\end{figure}

\quad\quad From Figure 9, we find that the energy extracting
efficiency in MCADAF is greater than that in ADAF for $a_ * > a_\ast
^{in} $, and the former could be 1.88 times at most than the latter
for $a_ * = 0.9$. This result implies that the luminosity could be
augmented significantly in MCADAF. As argued in Y06 the maximum
luminosity in LHAF is $L_{LHAF} \approx 16\% L_E $, then the maximum
luminosity in MCADAF could be $L_{MCADAF} \approx 30\% L_E $, which
could be used to explain the observation of the highest luminosity
of GX 339-4 in hard-state with $25 - 30\% L_E $ (Done {\&}
Gierlinski 2003; Zdziarski et al. 2004).

\quad\quad In this paper MCADAF is worked out by combining the MC
process with the self-similar solutions given in NY94. However,
there exists an obvious inconsistency in this model. The
self-similar solution is given based on pesudo-Newtonian potential
in NY94, while the MC is formulated in the context of general
relativity. Not long ago, some authors (Gammie and Popham 1998;
Popham and Gammie 1998) obtained the precise solutions in ADAF in
the context of general relativity, and these works provider a
possibility to improve MCADAF in our future work.

\noindent\textbf{Acknowledgments. }This work is supported by the
National Natural Science Foundation of China under Grant Numbers
10573006 and 10121503.

{}

\end{document}